\documentclass{article}

\usepackage{PRIMEarxiv}

\usepackage[utf8]{inputenc} % allow utf-8 input
\usepackage[T1]{fontenc}    % use 8-bit T1 fonts
\usepackage{hyperref}       % hyperlinks
\usepackage{url}            % simple URL typesetting
\usepackage{booktabs}       % professional-quality tables
\usepackage{amsfonts}       % blackboard math symbols
\usepackage{amsmath}
\usepackage{amssymb}
\usepackage{amsthm}
\usepackage{float}
\usepackage{placeins}
\usepackage{nicefrac}       % compact symbols for 1/2, etc.
\usepackage{microtype}      % microtypography
\usepackage{lipsum}
\usepackage{fancyhdr}       % header
\usepackage{graphicx}       % graphics
\graphicspath{{media/}}     % organize your images and other figures under media/ folder

\newtheorem{defn}{Definition}

\newtheorem{theorem}{Theorem}

\title{A review of cryptosystems based on multi layer chaotic mappings
\thanks{\textit{\underline{Citation}}: 
\textbf{Authors. Title. Pages.... DOI:000000/11111.}} 
}

\author{
  Awnon Bhowmik\\
  Department of Mathematics\\
  University of Central Florida\\
  \texttt{awnon.bhowmik@ucf.edu}\\
  \And
  Emon Hossain\\
  Department of Mathematics\\
  University of Dhaka\\
  \texttt{emon-2017613811@math.du.ac.bd}\\
  \And
  Mahmudul Hasan\\
  Department of Mathematics\\
  University of Dhaka\\
  \texttt{mahmudul-2016713604@math.du.ac.bd} \\
}

\begin{document}
\maketitle

\begin{abstract}
In recent years, a lot of research has gone into creating multi-layer chaotic mapping-based cryptosystems. Random-like behavior, a continuous broadband power spectrum, and a weak baseline condition dependency are all characteristics of chaotic systems. Chaos could be helpful in the three functional components of compression, encryption, and modulation in a digital communication system. To successfully use chaos theory in cryptography, chaotic maps must be built in such a way that the entropy they produce can provide the necessary confusion and diffusion. A chaotic map is used in the first layer of such cryptosystems to create confusion, and a second chaotic map is used in the second layer to create diffusion and create a ciphertext from a plaintext. A secret key generation mechanism and a key exchange method are frequently left out, and many researchers just assume that these essential components of any effective cryptosystem are always accessible. We review such cryptosystems by using a cryptosystem of our design, in which confusion in plaintext is created using Arnold's Cat Map, and logistic mapping is employed to create sufficient dispersion and ultimately get a matching ciphertext. We also address the development of key exchange protocols and secret key schemes for these cryptosystems, as well as the possible outcomes of using cryptanalysis techniques on such a system.
\end{abstract}

% keywords can be removed
% \keywords{First keyword \and Second keyword \and More}

\section{Introduction}
Protection from harmful sources has become an integral element of the discovery or creation process. For hiding very sensitive data, a variety of techniques of security are utilized, ranging from a basic authentication password to the most complicated Cryptography or Steganography or even quantum algorithms. Chaos signals are useful for picture encryption and enhance the resilience of cryptosystems against statistical assaults because of their key characteristics. Pseudorandom, non-periodicity, and high sensitivity to system parameters and initial states are some of these features. As a result, integrating cryptography with chaos theory is regarded as one of the most significant topics in information security \cite{abdullah2017image}. Chaos theory is also observed to be the underlying basis of \cite{romero2021image} and \cite{gao2021new}. Chaotic maps are basic functions that are easily iterated. As a result, chaos-based picture encryption systems are suitable for real-time applications.
\\~\\
Discrete and continuous temporal domains are allocated to chaotic maps. Iterated functions, which matched to rounds in cryptosystems, are the most common type of discrete maps. Chaotic cryptosystems are proposed based on the similarities between cryptography and discrete chaotic dynamic systems. Each map has certain parameters that are cryptographically comparable to encryption keys. A chaotic system is utilized to produce a pseudorandom key stream in stream ciphers, but the plaintext or secret key(s) are employed as the initial and control parameters in block ciphers. Finally, the chaotic systems are subjected to some repetition in order to retrieve the cipher-text. Cryptosystems have considerable challenges in terms of security and complexity. These should be taken into account while choosing a map and its parameters for cryptography \cite{shujun2001pseudo}.
\\~\\
Due to the current popularity of quantum computers and quantum technologies, even image encryption schemes have swung in that direction. Adaptive quantum logistic maps may be utilized to construct powerful cryptosystems, thanks to the popularity of machine learning and artificial intelligence techniques \cite{zaghloul2014color}. The healthcare industry is a vital sector where this sort of encryption protocol can be applied. For example, this article makes use of quantum image encryption using XOR operation \cite{abd2017robust}. The same author had published a paper \cite{abd2013new} where he combined random grids, error diffusion and chaotic permutation to encrypt secret images, and another paper where lifting wavelet transform is combined with quantum chaotic system for the purpose of image encryption \cite{abd2013new2}. Ron Rivest created the RC5 symmetric key block encryption technique in 1994. It is remarkable for being simple, quick (due to the use of just elementary computer operations such as XOR, shift, and so on), and memory efficient. A modified version of RC5 protocol can be used to encrypt images, as shown in \cite{amin2010efficient}. Numerous encryption protocols have been suggested over the recent years to improve security in the encryption of images. One such protocol is described in \cite{zhang2014diffusion}. All these research works simply states the fact that, if used appropriately, anything and everything can be used to make a proper trapdoor function. Chaotic maps are an attractive option for use in cryptographic systems due to their availability and initial condition sensitivity.

\section{Confusion and Diffusion}
Confusion is a cryptography method for generating faint cipher messages. The substitution algorithm allows for this method. In order to provide clarity, if one bit in the secret is changed, most or all bits in the cipher text will be changed as well. As a result of the confusion, the degree of ambiguity increases. Confusion is used in both stream and block ciphers. Confusion obscures the relationship between the cipher text and the key. The avalanche effect is a desirable characteristic of cryptographic algorithms, and it is commonly used in block ciphers and cryptographic hash functions \cite{feistel1973cryptography}. 
\\~\\
To produce mysterious plain messages, diffusion is employed. This is accomplished through the use of a transportation algorithm. If one bit in the plain text is changed, many or all of the elements in the encrypted matrix will be changed as well. As a result, redundancy is enhanced. Block ciphers are the only ones that use diffusion. Diffusion obscures the relationship between the encrypted text and the plain text.

\section{Logistic map}
The logistic map is a degree $2$ polynomial mapping (or recurrence relation) that is frequently used as a paradigmatic illustration of how complicated, chaotic behavior may develop from incredibly straightforward non-linear dynamical equations. The map was made famous in a $1976$ publication by the biologist Robert May \cite{may2004simple}, who used it as a discrete-time demographic model that was similar to Pierre François Verhulst's logistic equation \cite{mawhin2004legacy}. The logistic map is written mathematically.

\begin{equation}
\label{eq:logistic_map}
    x_{n+1}=rx_n(1-x_n)\qquad x_n\in(0,1),r\in(3.56995,4)
\end{equation}

\section{Arnold's Cat Mapping}
Arnold's cat map is a chaotic map from the torus into itself in mathematics. The Poincare Recurrence Theorem asserts that, given a particular degree of entropy, a limited, constrained, and ordered system will ultimately reach a state that is close to or precisely identical to the original state in some long but finite period \cite{furstenberg1981poincare}. This is true for certain dynamical systems, but not all. Arnold's cat map is one such example. It uses a specific matrix from the set of invertible $2\times 2$ matrices with integer components, which is closely connected to the well-known Fibonacci sequence. This ensures that the image's area is preserved.
\\~\\
Consider a square picture of $N \times N$ pixels, where each pixel's coordinates are represented by an ordered pair $(x, y)$ of real values in the range $[0, 1)$. Let the first iteration of Arnold's cat map be the matrix multiplication of all pixel coordinates by $A=\begin{pmatrix}1 & 1\\1 & 2\end{pmatrix}$. Then all values are taken modulo $1$, so that the generated coordinates are still in $[0, 1)$.
\\~\\
Arnold's cat map creates a discrete-time dynamical system whose evolution is determined by mapping iterations. It is symbolically written as
\begin{equation}
    \Gamma\left(\begin{array}{c}
        x_{n+1}\\
        y_{n+1}
    \end{array}\right)=\begin{pmatrix}1 & 1\\1 & 2\end{pmatrix}\begin{pmatrix}x_n\\y_n\end{pmatrix}\pmod{1}
\end{equation}The inverse of this mapping is given by
\begin{equation}
\label{eq:arnold_inv}
    \Gamma^{-1}\left(\begin{array}{c}
        x\\y
    \end{array}\right)=\begin{pmatrix}2 & -1\\-1 & 1\end{pmatrix}\begin{pmatrix}x\\y\end{pmatrix}\pmod 1
\end{equation}The Arnold cat map can be generalized by introducing two parameters $a,b$ and a constant $M$, giving us the following transformation

\begin{equation}
\label{gen_cat_map}
    \Gamma\left(\begin{array}{c}
        x_{n+1}\\
        y_{n+1}
    \end{array}\right)=\begin{pmatrix}1 & a\\b & 1+ab\end{pmatrix}\begin{pmatrix}x_n\\y_n\end{pmatrix}\pmod{M}
\end{equation}
We may formulate the following definition because the discrete-time dynamical system, i.e., the set of rules imposed by Arnold's cat map, will follow the Poincaré Recurrence Theorem and hence be periodic.

\begin{defn}
\normalfont
The \emph{minimal period} of Arnold’s discrete cat map is the smallest positive integer n such that $A_n\equiv \begin{bmatrix}1 & 0\\0 & 1\end{bmatrix} \pmod{N}$. It is denoted by $\prod_A(N)$.
\end{defn}

% \section{Henon Map}
%  The Henon map is a two-dimensional iterated discrete-time dynamical system with chaotic behavior. The Henon map is a two-dimensional iterated map with quadratic nonlinearity and chaotic solutions known as weird attractor. Strange attractors serve as a connection between chaos and fractals. Non-integer dimensions are common in strange attractors. Henon's attractor is a non-integer dimension attractor. The Hénon map gives the strange attractor with a fractal structure \cite{sonis1996once}.
% \\~\\
% The Henon map was presented as a simplified model of the Poincare map, which emerges from the solution of the Lorenz equations. It is symbolically represented by the following pair of recursive equations.

% \begin{equation}
% \label{eq:henon1}
%     \begin{cases}
%     x_{n+1}&=1-ax_n^2+by_n\\
%     y_{n+1}&=bx_n
%     \end{cases}
% \end{equation}
% where $a,b$ are positive bifurcation parameters. Notice that $y_n=x_{n-1}$. This allows us to write equation \ref{eq:henon1} as a single equation with two time delays \cite{sprott2006high}.
% \begin{equation}
%     \label{eq:henon2}
%     x_{n+1}=1-ax_n^2+bx_{n-1}
% \end{equation}

\section{Drawbacks of aforementioned chaotic maps}
\subsection{Logistic Map}
\subsubsection{Control Parameter Value Dependency}
The classical Logistic Map shows chaotic behavior for initial parameter values as specified in \ref{eq:logistic_map}. Although a necessary condition for equation \ref{eq:logistic_map} to demonstrate chaotic nature is that the value of the control parameter $r$ is greater than $3.57$, it is indeed not sufficient. The computation of the Lyapunov exponent for $r\in(3.57, 4)$ shows that there are a significant number of values of $r$ in this interval for which the Lyapunov exponent for the map is negative. 

\begin{figure}
    \centering
    \includegraphics[width=\textwidth]{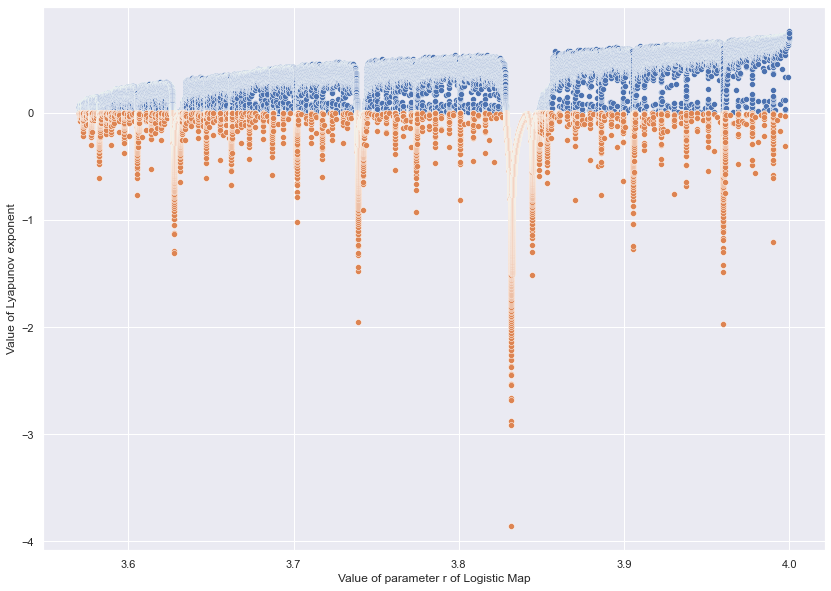}
    \caption{Lyapunov Exponent for Logistic map}
    \label{LE_Logistic}
\end{figure}

This implies that there are a significant number of values of $r$ for which \ref{eq:logistic_map} will not produce enough chaoticity and as a result, any cryptosystem utilizing the logistic map will not generate enough random-like behavior during the encryption process. Thus the strength of the encryption largely depends on whether a proper control or initial parameter is chosen. Furthermore, the statistical complexity of the logistic map is strongly related to the choice of the control parameter \cite{arroyo2008inadequacy} and many control parameter values show a low statistical complexity making them more vulnerable to attacks compared to other choices.

\subsubsection{Complexities in Secret Key Generation}
Any practically implemented cryptosystem will have to utilize a secret key generation process and a key exchange process. We have noticed while implementing a logistic map-based system that, several complexities arise while trying to implement these two processes. This complexity stems directly from the fact that how dependent the chaotic behavior of the logistic map is even when choosing the control parameter value in the interval $(3.57, 4)$. We can't implement a secret key generation process based on the assumption that the logistic map will show proper chaotic behavior for all values in this interval. Furthermore, suppose we decide to generate secret keys based on entropy measurement tools such as the Lyapunov exponent. In that case, further complexities may arise, as only one entropy measure might not be enough to ensure proper chaotic behavior from parameters generated from a secret key in the case of logistic map \cite{pastor1997revision}. This makes ensuring the cryptographic strength of a generated secret key for the logistic map very computationally expensive.

\subsubsection{Infeasibility of Determining the Key Space}
A properly implemented cryptosystem must have a well defined key space of keys that are cryptographically strong and contains no equivalent keys, i.e., every key in the key space is unique. In case of logistic map, it is not possible to define the underlying key space in a clear and concise manner. Due to the existence of infinitely many values of the parameter $r$ in the interval $(3.57, 4)$ for which the logistic map may not produce chaotic behavior, any implemented secret key generation algorithm will produce a significant number of secret keys that will not actually be cryptographically strong and will not produce enough random-like behavior in the generated ciphertext. It is not only computationally infeasible to ensure that every secret key generated through an algorithm will produce effective control parameters, but also to our knowledge there are no known mathematical methods of determining continuous sub-intervals of $(3.57, 4)$ for which the logistic map ensures enough chaotic behavior.  

\subsection{Arnold's Cat Map}
\subsubsection{Periodicity of Arnold's Cat Map}
The periodicity of Arnold's Cat Map was studied by Dyson et al. where they mathematically showed that the period of the map has an upper bound and a lower bound for any matrix of dimension $N\times N$\cite{dyson1992period}. For an image of $N\times N$ pixels it was shown that the minimal period or the minimum number of iterations required for the image to return to its original state never exceeds $3N$. Furthermore for special cases this upper bound for the minimal period to occur can be further reduced as indicated by the following theorem.   

\begin{theorem}
\normalfont
\begin{equation}
\label{eq.1}
    m_N\le 3N
\end{equation}
Moreover, equation $\ref{eq.1}$ holds with equality iff
\begin{equation}
\label{eq.2}
    N=2\cdot 5^y
\end{equation}
For all $N$ except for equation \ref{eq.2} we have
\begin{equation}
\label{eq.3}
    m_N\le 2N
\end{equation}
with equality only for
\begin{equation}
\label{eq.4}
    N=5^y\qquad N=6\cdot 5^y
\end{equation}
For all $N$ except for equation \ref{eq.2} and \ref{eq.4} we have
\begin{equation}
    \label{eq.5}
    m_N\le\dfrac{12}{7}N
\end{equation}
\end{theorem}

Bao and Yang \cite{bao2012period} studied both the Discrete Arnold's Cat Map and General Cat Map expanding on the work of Dyson et al. developing further formulae, theorems and algorithm to determine the minimal period for both cases, although based on some restrictions. 
\subsubsection{Fibonacci Sequence and Periodicity of Arnold's Cat Map}
There is a strong relationship between the minimal period of Arnold's Cat Map and the Fibonacci sequence. 

\begin{defn}
\normalfont
The Fibonacci sequence is defined by the recursive relation $F_n=F_{n-1}+F_{n-2}$ with $F_0=0$ and $F_1=1$.
\end{defn}
\begin{flushleft}
Using this relation, the first few Fibonacci numbers are $$0,1,1,2,3,5,8,13,21,34,55,89,\cdots$$

The Fibonacci sequence may be found in a variety of places, ranging from Pascal's triangle \cite{green1968recurrent} to real-life items like a pineapple shell. Powers of the matrix $F=\begin{pmatrix}0 & 1\\1 & 1\end{pmatrix}=\begin{pmatrix}F_0 & F_1\\F_1 & F_2\end{pmatrix}$ will generate numbers of the Fibonacci sequence $$F^n=\begin{pmatrix}F_{n-1} & F_n\\F_n & F_{n+1}\end{pmatrix}$$
setting $n=2$ gives $$F^2=\begin{pmatrix}0 & 1\\1 & 1\end{pmatrix}\begin{pmatrix}0 & 1\\1 & 1\end{pmatrix}=\begin{pmatrix}1 & 1\\1 & 2\end{pmatrix}=A$$
The Fibonacci numbers are also seen when we take powers of the matrix $A$. Namely, $$A^n=\begin{pmatrix}1 & 1\\1 & 2\end{pmatrix}^n=\begin{pmatrix}F_{2n-1} & F_{2n}\\F_{2n} & F_{2n+1}\end{pmatrix}$$
\\
From the definition of the minimal period of Arnold's cat map, we know that we seek the smallest $n$ such that $$A^n\equiv I \pmod N$$, where $I$ is the identity matrix. This consequently implies that we require the smallest $n$ such that $$F_{2n-1}\equiv 1 \pmod N$$ and $$F_{2n}\equiv 0 \pmod N$$
\end{flushleft}

\subsubsection{Cryptanalytic Attacks on Cryptosystems Utilizing Arnold's Cat Map}
Based on the findings of Dyson et al. the classical cat map, i.e., Arnold's Cat Map's period of recurring to the initial state always lies in a bounded range of integer values making it insecure when either a naive brute force attack is attempted or an attack mathematically inspired is implemented. If a cryptosystem utilizes Arnold's Cat Map on an image of $N\times N$ pixels, a brute force attack will always be sufficient to determine the original state of an image by iterating the map and visually observing the resultant image. As the cat map is most often used to introduce confusion in the plaintext when designing a cryptosystem based on multi layer chaotic maps, the subsequent chaotic map that will change the values of the pixels in the image will have to be very cryptographically secure and well designed, otherwise the whole cryptosystem will be vulnerable.

% \subsection{Henon Map}
% For $b=0.3$ and $a\in[1.06,1.4]$, the Henon map is in a chaotic state. The classical Hénon map has the disadvantages of small parameter range and low complexity \cite{li2022two}.

\section{Probable Improvements}
\subsection{Logistic Map}
\subsubsection{Generalized Logistic Map}
A modification of the Logistic Map to improve the underlying key space has been suggested\cite{lawnik2017generalized}. This improvement mainly aims to substantially increase the amount of parameter values that generates the intended chaoticity of the classical logistic map. The generalized logistic map (GLM) is given by the formula

\begin{equation}
    x_{n+1}=\begin{cases}-\dfrac{q}{p^2}(p-x_n)^2+q & 0\le x_n\le p\\ -\dfrac{q}{(1-p)^2}(p-x_n)^2+q & p<x_n\le 1\end{cases}
\end{equation} where $p\in(0,1)$ and $q\in[0,1]$.

Although GLM may increase the size of the key space and give us more room to improve secret key generation techniques, the underlying problem of the classical logistic map is still not overcome. As GLM also increases total number of parameter values of low statistical complexity and parameter values with a negative Lyapunov exponent. This in the end doesn't solve the issues regarding the complexity of choosing good parameter values, implementing proper secret key generation methods or complexities related to defining the key space. But if logistic map is to be used, better results may be obtained if it is substituted with GLM.

\subsection{Arnold's Cat map}
\subsubsection{Generalized Cat Map}
The generalized cat map as mentioned in \ref{gen_cat_map} offers more security features in contrast to Arnold's Cat Map. By introducing two parameters $a$ and $b$, determining the periodicity of the map can be made sufficiently complex. Bao et al. developed formulae to derive the minimal period of the general cat map for specific values of the parameters $a$ and $b$, but no fixed upper or lower bound for the recurrence of the original matrix was found. 

\begin{theorem}
\normalfont
For the generalized  Cat Map, the following conditions hold:
\begin{enumerate}
    \item If $N=ab+1$ with $a\neq 1$ and $b\neq 1$, then $\Gamma(N)=6$
    \item If $N=ab+2$, then $\Gamma(N)=4$
    \item If $N=ab+3$, then $\Gamma(N)=3$
    \item If $N=a^2b^2+5ab+5$, then $\Gamma(N)=5$
    \item If $N=a^3b^3+7a^2b^2+14ab+7$, then $\Gamma(N)=7$
    \item If $N=a^2b^2+4ab+2$, then $\Gamma(N)=8$
\end{enumerate}
\end{theorem}

This makes determining the minimal period of the generalized cat map less susceptible to brute force attacks and due to Bao et al. several precautions can be taken while developing a cryptosystem based on the generalized cat map to ensure that the values of the parameters $a$ and $b$ are chosen in a way so that the system is not vulnerable to specific cryptanalytic attacks based on mathematical theory. 

\section{Implemented System Drawbacks}
\subsection{Secret Key}
We leverage the Secrets module, a part of Python's standard library, to produce a cryptographically strong random string of $40$ characters consisting of hexadecimal digits, i.e, 0-9 and A-F. This random string will be used as the secret key of our cryptosystem.

\subsection{Generating Parameters from the Secret Key}
Our proposed encryption algorithm uses four parameters during the whole encryption process. All four of these parameters will be extracted from the secret key based on the following workflow:
\begin{enumerate}
    \item Split the secret key of $40$ characters into four equal chunks of 10 characters.
    \item Take the decimal values of the characters of the first chunk and append the values to 3 as decimal point digits. For example, if the values are evaluated to be $7625674573$, we append it to $3$ to get $3.7625674573$. If by chance the first digit is less than $6$ we change the first digit to $6$ to ensure the value falls in the interval $(3.6, 4)$. In this way, we produce our first parameter $r$.
    \item In a similar manner we produce the second parameter $x_0$ from the second chunk.
    \item The value of the base used during prepossessing of the data is also extracted from the third chunk following the process detailed in Step 2, but in this case a decimal point is appended after the first value. For example, if we get $1323445745$, we convert it to $1.323445745$.
    \item The iteration number used to scramble matrices using Arnold's Cat Mapping is taken to be the decimal value of the middle hexadecimal character of the fourth chunk.
\end{enumerate}

\subsection{Key Space and Basic Features}
Our proposed secret key scheme and the subsequent generation of parameters from it shows strong avalanche effect, as changing even a single character from the first $30$ characters will change at least one parameter. In this case we may choose other values for the symbols A-F and make the secret key generation process more difficult to attack.
\\~\\
Although, the availability of $40$ characters imply that the initial key space is of size less than or equal to $2^{160}$, the use of the last chunk is not as potent as the first $3$ chunks, even then the key space is approximately close to $2^{120}$. As a basic requirement for a secret key of a cryptosystem is that it's key space should at least be greater than $2^{100}$, it follows that our secret key space fulfills this requirement. We also investigated the possibility of acquiring equivalent keys, but to our knowledge it is not possible to generate equivalent keys based on this scheme, as the value of each index of the first $30$ characters matters in generation of the parameters.
\\~\\
Furthermore, we kept our secret key generation scheme and extraction of parameters from it as simple as possible so that it does not affect the overall performance of our algorithm. Hence, we have intentionally avoided the use of any mathematical computation during the parameter extraction process.

\subsection{Complexities in Actual Key Space Size Due to Logistic Map}
The effective key space of our implemented secret key generation algorithm should be greater than $2^{100}$, but as we have noticed while reviewing the logistic map, this will not be the case. We summarize the complexities below:
\begin{enumerate}
    \item Due to the availability of an infinite number of points for which $r\in(3.6, 4)$ gives negative values for the Lyapunov exponent, it is not possible to determine how many points in the key space are actually viable for cryptographic purposes.
    \item We can also notice that for different values of $x_0\in(0,1)$ the values of $r$ which gives positive values for the Lyapunov exponent varies. 
    % \item If our secret key generation algorithm is based on one or more entropy measure tools such as Lyapunov exponent, it becomes computationally expensive to ensure the security of the secret keys and often these measures might not produce accurate results.
    \item It becomes computationally expensive to assure the security of the secret keys if our secret key generation technique is dependent on one or more entropy measure tools, such as the Lyapunov exponent, and frequently these measures might not yield reliable findings.
    \item The above points makes creating a proper secret key generation algorithm with a properly defined key space infeasible for logistic map.   
\end{enumerate}
%%%%%%%%%%%%%
% the following section was created with dummy data
% real data will be generated soon and replaced with that
% check the latex format and the text to improve it 
%%%%%%%%%%%%%
\subsection{Algorithm}
The algorithm of the implemented system are explained below.
\subsubsection{Encryption}
\label{enc}
\begin{enumerate}
    \item Input text of arbitrary string length. Create an array from ASCII values of the string. Apply logarithm function with arbitrary base $b$. Generate $\left(\dfrac{N}{2}\right)^2-\dfrac{N}{2}$ $\Big($ or $\left(\dfrac{N+1}{2}\right)^2-\dfrac{N+1}{2} \Big)$ floating numbers depending on $N$ is even or odd respectively from uniform distribution between $(\log_b 65,\log_b 122)$ and turn all of these values into a $\dfrac{N}{2}\times\dfrac{N}{2}$ or $\Big(\dfrac{N+1}{2}\times\dfrac{N+1}{2}\Big)$ square random matrix. 
    \item Apply Arnold's Cat map, making sure that there is a significant difference between original and final matrix.
    \item Choose private keys $r$, initial $x_0$ to generate a $\dfrac{N}{2} \times \dfrac{N}{2}$ $\Big($ or $\left(\dfrac{N+1}{2}\times \dfrac{N+1}{2} \right)\Big)$ size square noise matrix from Logistic map.
    \item Apply XOR between the two matrix. Return the encrypted random matrix.
\end{enumerate}
This ends the encryption algorithm.

\subsubsection{Decryption}
\label{dec}
\begin{enumerate}
    \item Use the private key parameters to regenerate the noise matrix from Logistic map.
    \item Apply XOR operator between the noise matrix and the encrypted matrix.
    \item Use the private key parameters to reverse the Arnold cat map effect from the encrypted matrix. This returns the preprocessed $\dfrac{N}{2} \times \dfrac{N}{2}$ $\Big($ or $\left(\dfrac{N+1}{2}\times \dfrac{N+1}{2} \right)\Big)$ square matrix (undo the shuffling). This can be done by using the inverse Arnold Cat mapping.
    \item Extract $N$ elements from $\left(\dfrac{N}{2}\right)^2$ $\Big($or $\left(\dfrac{N+1}{2}\right)^2\Big)$ elements of that random matrix.
    \item Convert the resultant array with $(\text{ASCII values})$ into the original text.
\end{enumerate}
This ends the decryption algorithm.

\subsection{Statistical Analysis \& Complexities}
\subsubsection{Correlation  Analysis}

The  digital  multimedia  data  has  high  correlations  among  their  adjacent  pixels/frames.  An encryption algorithm should be able to eliminate the correlations among the vertically, horizontally and diagonally adjacent pixels in images. Highly correlated pixels have a value of coefficient as ±1, where as,  the  uncorrelated  pixels  have  correlation coefficient  close  to  0.  Pixels correlations  in  original  and other encrypted images (lena) for three directions are provided in Table \ref{tab:sensitivity_analysis}. 

% comparison for lena image 
% copy the citation from the below paper -> Awnon bhai  https://www.researchgate.net/publication/260947149_An_Inter-Component_Pixels_Permutation_Based_Color_Image_Encryption_Using_Hyper-chaos
% I already generate the data

\begin{table}[hbt!]
    \centering
    \begin{tabular}{lccl}
        \hline
        \textbf{Method} & \textbf{Horizontal} & \textbf{Vertical} & \textbf{Diagonal}\\\hline
        Original & 0.9597 & 0.9792 & 0.9570 \\
        Rhouma et al. \cite{rhouma2009ocml} & 0.0845 & 0.0681 & NA\\
        Hongjun et al. \cite{liu2010color} & 0.0965 & -0.0318 & 0.0362 \\
        Huang \cite{huang2009multi} & 0.1257 & 0.0581 & 0.0226 \\
        Proposed & 0.0297 & -0.3665& -0.2999\\\hline
    \end{tabular}
    \caption{Sensitivity Analysis}
    \label{tab:sensitivity_analysis}
\end{table}

\subsubsection{Entropy Analysis}
Entropy is a statistical parameter that is defined to measure the uncertainly and randomness of a bundle of data. According to Shannon theory, image entropy is the number of bits that are necessary to encode every pixel of the image. The optimal value for entropy of an encrypted image is ~8. This quantity describes the random pattern and texture of pixels in an encrypted image and is calculated by:

\begin{equation}
    H=-\sum_{i=0}^{n-1}p_i\log_b p_i
\end{equation} where $n$ is the total number of gray levels (i.e., 256) and $P_i$ is the probability of incidence of intensity $i$ in the current image. $P_i$ is the number of pixels with intensity $i$ divided by the total number of pixels. The base-2 logarithm will present the calculated entropy in bits. A sample test run was performed using different strings with varying string lengths demonstrate entropy before and after encryption using \href{https://scikit-image.org/docs/stable/api/skimage.measure.html#skimage.measure.shannon_entropy}{shannon entropy}. Table \ref{tab:entropy_analysis} shows the outcome.

\FloatBarrier
\begin{table}[hbt!]
\centering
\begin{tabular}{lllll}
    \hline
     \textbf{String/Image} & 	\textbf{Before} &	\textbf{After}\\\hline
\begin{tabular}[c]{@{}l@{}}The quick brown fox\\ jumps over the lazy dog\end{tabular}  & 4.4877  & 8.9188\\ \hline
Lorem Ipsum is simply dummy text                                                      	& 3.7508 & 8.0\\ \hline
\begin{tabular}[c]{@{}l@{}}In cryptography, encryption\\ is the process of encoding information\end{tabular}  & 4.04918 & 10.0887\\ \hline
\begin{tabular}[c]{@{}l@{}} lena\_gray\_512 \end{tabular}  & 7.44506 & 17.9999\\ \hline

\end{tabular}
\caption{Entropy analysis}
\label{tab:entropy_analysis}
\end{table}
\FloatBarrier

\subsection{Differential Analysis}
Table \ref{tab:diff_analysis} displays the results of several tests achieved during differential analysis.

\FloatBarrier
\begin{table}[hbt!]
\centering
\begin{tabular}{lcccccc}
    \hline
     \textbf{Image} & 	\textbf{NCPR}	& \textbf{UACI} &	\textbf{MSE}&\textbf{PSNR}\\\hline
	\begin{tabular}[c]{@{}l@{}}lena\_gray\_512\end{tabular}& 99.5735 & 50.9174& 135.0196& 26.8268                   \\ \hline
\begin{tabular}[c]{@{}l@{}}boat\_gray\_512\end{tabular} & 99.6143& 49.2415& 134.7697& 26.8348 \\ \hline
\begin{tabular}[c]{@{}l@{}}mandril\_gray\_512 \end{tabular}      & 99.5830 & 51.3193& 135.1067& 26.8240  \\ \hline

\end{tabular}
\caption{Differential analysis}
\label{tab:diff_analysis}
\end{table}
\FloatBarrier

\section{Conclusion}
In this brief review endeavor, we have mentioned various drawbacks of numerous chaotic maps, making them insecure for use in a modern cryptosystem design. A multi layer cryptosystem based on these chaotic maps may increase the computational complexity of the program but not the actual strength of an encryption protocol. Hence such designs are not recommended to be used in further cryptosystem designs, especially without any modifications that improve their security features. There are however, other chaotic maps which do not yet exhibit such drawbacks and is neither vulnerable to brute force attacks enabled by modern hardware capabilities nor susceptible to various known and widely used cryptanalytic attacks. These improved and more secure chaotic maps may be used for designing new chaos-based cryptosystems. Yet as we have seen while reviewing the logistic map, the control parameter values of these new maps should be monitored carefully to ensure that sufficient chaotic behavior from them is actually achieved. 

% \section*{Acknowledgments}
% This was was supported in part by......

%Bibliography
\bibliographystyle{unsrt}  
\bibliography{references}

\end{document}